\documentclass[twocolumn,showkeys,aps,prb,showpacs]{revtex4-1}
\usepackage{graphicx}
\usepackage[CJKbookmarks,dvipdfm,colorlinks,linkcolor=blue,citecolor=blue]{hyperref}

\begin{document}

\title{Potential thermoelectric materials $\mathrm{CsMI_3}$ (M=Sn and Pb) in  perovskite structures from the first-principles calculations}

\author{San-Dong Guo and Jian-Li Wang}
\affiliation{Department of Physics, School of Sciences, China University of Mining and
Technology, Xuzhou 221116, Jiangsu, China}
\begin{abstract}
The thermoelectric properties of halide perovskites $\mathrm{CsMI_3}$ (M=Sn and Pb) are investigated from  a combination of  first-principles calculations and semiclassical Boltzmann transport theory by  considering both the electron and phonon transport. The electronic part is performed using a modified Becke and Johnson (mBJ) exchange potential, including spin-orbit coupling (SOC), while the phonon part is computed using generalized gradient approximation (GGA).
It is found that  SOC has  remarkable detrimental effect on n-type  power factor, while has a negligible influence in p-type doping, which can be explained  by considering SOC effect on conduction and valence bands. Calculated results show  exceptionally low lattice thermal conductivities in $\mathrm{CsSnI_3}$ and $\mathrm{CsPbI_3}$, and the corresponding room-temperature lattice thermal conductivity is 0.54   $\mathrm{W m^{-1} K^{-1}}$ and 0.25 $\mathrm{W m^{-1} K^{-1}}$. At 1000 K, the maximal figure of merit $ZT$ is up to 0.63 and 0.64 for $\mathrm{CsSnI_3}$ and $\mathrm{CsPbI_3}$ with scattering time $\tau$=$10^{-14}$ s, and the peak  $ZT$ is 0.49 and 0.41 with  $\tau$=$10^{-15}$ s. These results  make us believe that $\mathrm{CsMI_3}$ (M=Sn and Pb) in  perovskite structures may be potential thermoelectric materials.
\end{abstract}


\maketitle

\section{Introduction}
Thermoelectric materials enable the direct conversion from heat to electricity, which may make  essential contributions
to the crisis of energy\cite{s1,s2}. A good thermoelectric material should have high dimensionless  figure of merit $ZT=S^2\sigma T/(\kappa_e+\kappa_L)$, where S, $\sigma$, T, $\kappa_e$ and $\kappa_L$ are the Seebeck coefficient, electrical conductivity, absolute  temperature, the electronic and lattice thermal conductivities, respectively. A high $ZT$ material requires high power factor ($S^2\sigma$) and low thermal conductivity ($\kappa=\kappa_e+\kappa_L$). They  generally influence each other, and a counteracted relationship between electrical conductivity and thermal conductivity or Seebeck coefficient  is often found.
The excellent classic thermoelectric materials include bismuth-tellurium systems\cite{s3,s4}, silicon-germanium alloys\cite{s5,s6}, lead chalcogenides\cite{s7,s8} and skutterudites\cite{s9,s10}. For  thermoelectric research,  searching for potential high $ZT$ materials with classic common lattice structure, like perovskite structure,  is interesting and challenging.

The cubic perovskite structure  $\mathrm{SrTiO_3}$  has attracted growing attention for thermoelectric power generation\cite{s17}. The $ZT$ value of undoped  $\mathrm{SrTiO_3}$  is less than 0.5 due to its high thermal conductivity\cite{s18}, which can be reduced by introducing point defects\cite{s19,s20}.  Searching other perovskites for more efficient thermoelectric applications is imperative and amusing.
Hybrid  $\mathrm{AMX_3}$ perovskites (A=Cs,  $\mathrm{CH_3NH_3}$; M=Sn, Pb; X=halide) have recently attracted a great
deal of attention for solar cell designs, which can realize up to 15\% energy conversion efficiencies\cite{s11,s12,s13,s14}.
A fair amount of theoretical  works have been performed to investigate their electronic structures, phonons and optical properties\cite{t1,t2,t3,t4,t5,t6}.
The $\mathrm{CsMI_3}$ (M=Sn and Pb)  of them  under reasonable hydrostatic pressure can turn into three-dimensional topological insulators, which has been predicted by first-principles calculations\cite{s15,s16}.

Here, we report on the thermoelectric properties of cubic  $\mathrm{CsMI_3}$ (M=Sn and Pb) in  perovskite structures from  a combination of  first-principles calculations and semiclassical Boltzmann transport theory.
The SOC can produce obvious effects on power factor of many thermoelectric materials\cite{so1,so2,so3,so4,so5,gsd3,gsd4,so6}, so the SOC  is included  in our calculations of electronic part. Calculated results show SOC has  a noteworthy reduced influence for n-type power factor.  It is noteworthy that ultralow lattice thermal conductivities are attained for $\mathrm{CsMI_3}$ (M=Sn and Pb), and  the corresponding  lattice thermal conductivity  at 300 K is 0.54   $\mathrm{W m^{-1} K^{-1}}$ and 0.25 $\mathrm{W m^{-1} K^{-1}}$, which is compared with lattice thermal conductivity of 0.23  $\mathrm{W m^{-1} K^{-1}}$ in SnSe with an unprecedented $ZT$ of 2.6 at 923 K\cite{zhao}. Finally, the dimensionless thermoelectric figure of merit $ZT$ is calculated with $\tau$=$10^{-14}$ s or $\tau$=$10^{-15}$ s , and can be up
to about  0.6 or 0.4 at high temperature by the optimized doping.

\begin{figure}
  \includegraphics[width=5cm]{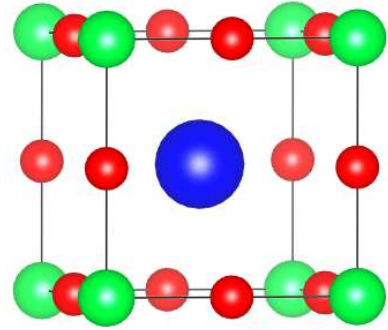}
  \caption{(Color online) The crystal structure of $\mathrm{CsMI_3}$ (M=Sn and Pb). The largest blue ball represent Cs atom, the medium green balls X, and the smallest red balls I.}\label{st}
\end{figure}

The rest of the paper is organized as follows. In the next section, we shall
describe computational details. In the third section, we shall present the electronic structures and  thermoelectric properties of  $\mathrm{CsMI_3}$ (M=Sn and Pb). Finally, we shall give our discussions and conclusion in the fourth
section.
\begin{figure}
  \includegraphics[width=8.0cm]{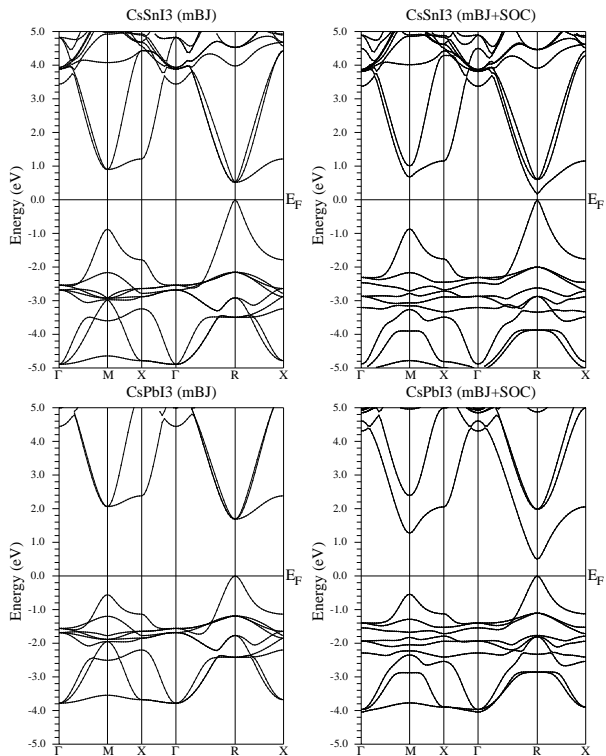}
  \caption{The energy band structures of $\mathrm{CsMI_3}$ (M=Sn and Pb)  using mBJ (Left) and mBJ+SOC (Right).}\label{band}
\end{figure}

\begin{figure*}
  \includegraphics[width=15cm]{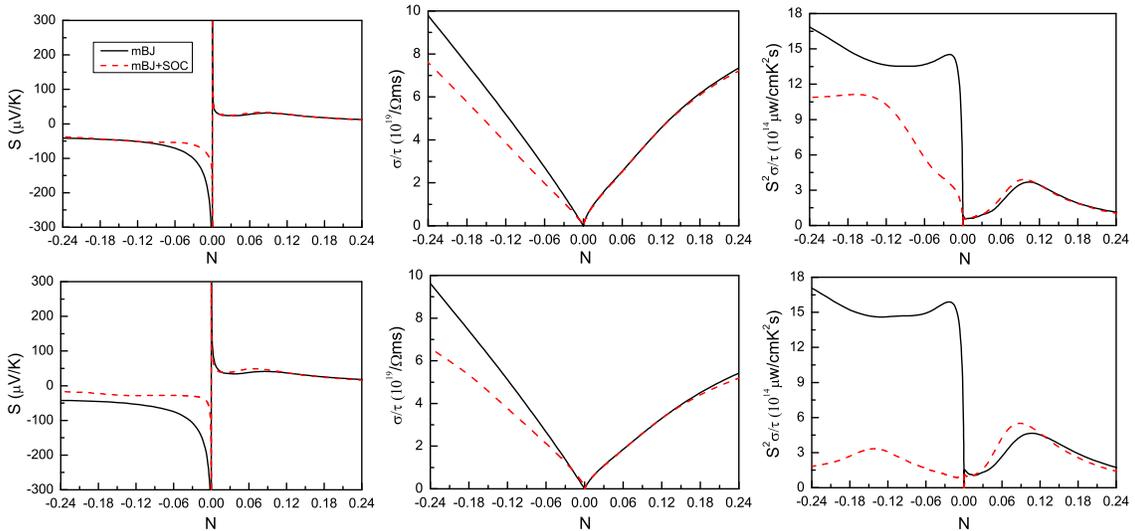}
  \caption{(Color online)  At room temperature,  transport coefficients of $\mathrm{CsSnI_3}$ (Top panel) and $\mathrm{CsPbI_3}$ (Bottom panel) as a function of doping level (N):  Seebeck coefficient S, electrical conductivity with respect to scattering time  $\mathrm{\sigma/\tau}$  and   power factor with respect to scattering time $\mathrm{S^2\sigma/\tau}$   calculated with mBJ (Black solid lines) and mBJ+SOC (Red dash lines). The N means electrons (minus value) or holes (positive value) per unit cell.}\label{t1}
\end{figure*}
\begin{figure*}
  \includegraphics[width=14cm]{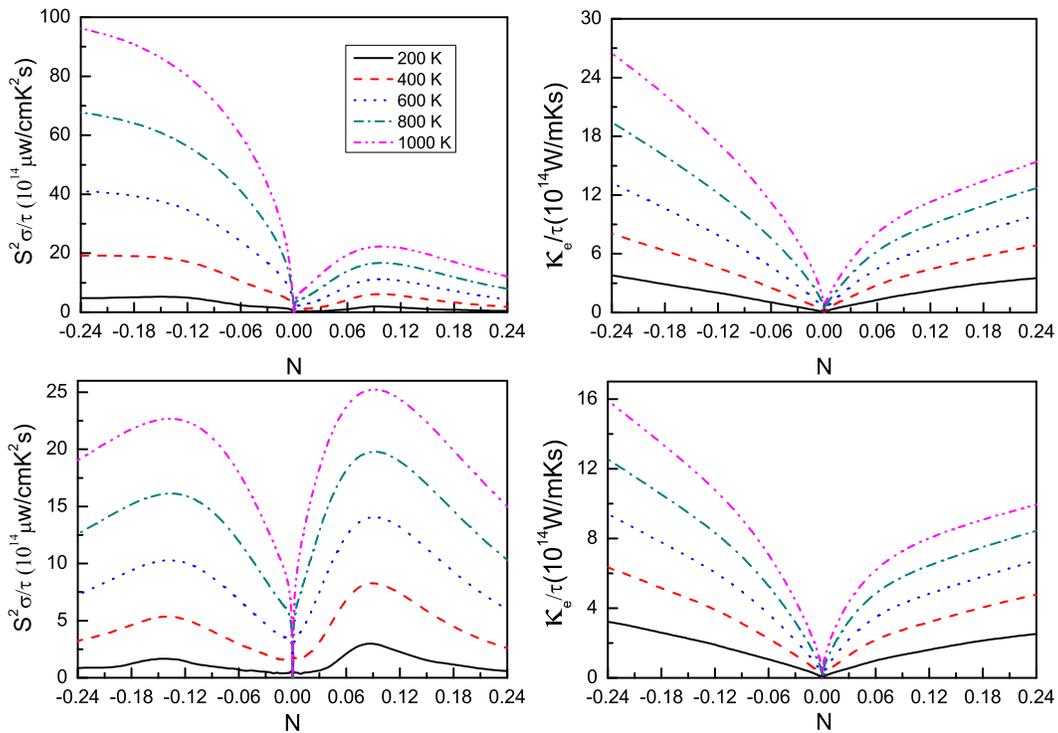}
  \caption{(Color online) The power factor with respect to scattering time  $\mathrm{S^2\sigma/\tau}$ and  electronic thermal conductivity with respect to scattering time $\mathrm{\kappa_e/\tau}$  of $\mathrm{CsSnI_3}$ (Top panel) and $\mathrm{CsPbI_3}$ (Bottom panel) as a function of doping level with temperature  being 200, 400, 600 , 800 and 1000 (unit: K) using mBJ+SOC. }\label{t2}
\end{figure*}

\section{Computational detail}
The electronic structures of $\mathrm{CsMI_3}$ (M=Sn and Pb) are performed using
a full-potential linearized augmented-plane-waves method
within the density functional theory (DFT) \cite{1}, as implemented in
the WIEN2k  package\cite{2}.  We employ Tran and Blaha's mBJ
 exchange potential plus local-density approximation (LDA)
correlation potential  for the
exchange-correlation potential \cite{4}, which has been known to produce
more accurate band gaps than LDA and GGA.
 The SOC was included self-consistently \cite{10,11,12,so} due to containing heavy elements, which leads to band splitting, and produces important effects on power factor. We use 5000 k-points in the
first Brillouin zone for the self-consistent calculation,  make harmonic expansion up to $\mathrm{l_{max} =10}$ in each of the atomic spheres, and set $\mathrm{R_{mt}*k_{max} = 8}$. The self-consistent calculations are
considered to be converged when the integration of the absolute
charge-density difference between the input and output electron
density is less than $0.0001|e|$ per formula unit, where $e$ is
the electron charge. Transport calculations, including Seebeck coefficient, electrical conductivity and electronic
 thermal conductivity,
are performed through solving Boltzmann
transport equations within the constant
scattering time approximation (CSTA) as implemented in
BoltzTrap\cite{b}, and reliable results  have
been obtained for several materials\cite{b1,b2,b3}. To
obtain accurate transport coefficients, we use 140000 k-points in the
first Brillouin zone for the energy band calculation. The  lattice thermal conductivities are calculated within  the linearized phonon Boltzmann equation
by using Phono3py+VASP codes\cite{pv1,pv2,pv3,pv4}. For the third-order force constants, 2$\times$2$\times$2 supercells
are built, and reciprocal
spaces of the supercells are sampled by  3$\times$3$\times$3 meshes. To compute lattice thermal conductivities, the
reciprocal spaces of the primitive cells  are sampled using the 20$\times$20$\times$20 meshes.

\section{MAIN CALCULATED RESULTS AND ANALYSIS}
$\mathrm{CsMI_3}$ (M=Sn and Pb) belong to  perovskite semiconductors, which consists of a network
of corner-sharing $\mathrm{MI_6}$ octahedra, and the schematic crystal structure is shown in \autoref{st}.
Based on the experimental structures, the electronic structures of $\mathrm{CsMI_3}$ (M=Sn and Pb) are investigated using mBJ and mBJ+SOC, and  the  energy band structures are plotted in \autoref{band}. Both mBJ and mBJ+SOC show $\mathrm{CsMI_3}$ (M=Sn and Pb) are  direct-gap semiconductor, with the conduction band minimum (CBM) and valence band maximum (VBM) at the R point. The mBJ and mBJ+SOC energy band gap values are 0.52 eV (1.69 eV)  and 0.17 eV (0.50 eV) for $\mathrm{CsSnI_3}$ ($\mathrm{CsPbI_3}$), respectively.  The conduction bands are dominated by  M-6p states, and the CBM is threefold degenerate at the  absence of SOC.   The VBM  is a mixture of I-p and M-s states, which is nondegenerate. The SOC can remove  band degeneracy, and leads to a spin-orbital splitting value of 0.43 eV  and 1.49 eV for  $\mathrm{CsSnI_3}$ and  $\mathrm{CsPbI_3}$ at CBM.  It is clearly seen that, near the Fermi level, the SOC has  more obvious  influence on the conduction bands  than on the valence bands.
The related data are shown in \autoref{tab0}, and the mBJ gaps are larger than GGA or LDA ones, but are less than GW or HSE ones\cite{t1,t3,t4,t5,t6}.
\begin{table}[!htb]
\centering \caption{ The experimental lattice constant $a$  ($\mathrm{{\AA}}$); the calculated energy band gap values  with mBJ $E_1$ (eV) and mBJ+SOC $E_2$ (eV); $E_1-E_2$ (eV);  spin-orbit splitting $\Delta$ (eV)  at the CBM.}\label{tab0}
  \begin{tabular*}{0.48\textwidth}{@{\extracolsep{\fill}}cccccc}
  \hline\hline

 Name & $a$  & $E_1$ & $E_2$&$E_1-E_2$ &$\Delta$\\\hline\hline
 $\mathrm{CsSnI_3}$&6.22  & 0.52 &0.17&0.35&0.43\\\hline
 $\mathrm{CsPbI_3}$&6.29  &1.69 & 0.50& 1.19&1.49\\\hline\hline
\end{tabular*}
\end{table}
\begin{table*}[!htb]
\centering \caption{ Peak $ZT$ for both n- and p-type at 1000 K with $\tau$=$10^{-14}$ s and $\tau$=$10^{-15}$ s, and   the corresponding doping concentrations. The doping concentration equals  $\mathrm{4.16\times10^{21}cm^{-3}}$($\mathrm{4.02\times10^{21}cm^{-3}}$)$\times$ doping level for $\mathrm{CsSnI_3}$ ($\mathrm{CsPbI_3}$). }\label{tab}
  \begin{tabular*}{0.96\textwidth}{@{\extracolsep{\fill}}ccccccccc}
  \hline\hline

                 &    & $\tau$=$10^{-14}$ s & & &   &  $\tau$=$10^{-15}$ s& &\\\hline
                   &n&&p&&n&&p&\\
  Name&($\mathrm{\times10^{19}cm^{-3}}$)&$ZT$&($\mathrm{\times10^{19}cm^{-3}}$)&$ZT$&($\mathrm{\times10^{19}cm^{-3}}$)&$ZT$&($\mathrm{\times10^{19}cm^{-3}}$)&$ZT$\\\hline\hline
 $\mathrm{CsSnI_3}$&4.16&0.63&1.08&0.36&9.78&0.49&1.88&0.19\\
 $\mathrm{CsPbI_3}$&0.53&0.64&0.60&0.65&1.14&0.38&1.53&0.41\\\hline\hline
\end{tabular*}
\end{table*}

Next, we calculate  semi-classic transport coefficients  using CSTA Boltzmann theory.
The rigid band approach is used to mimic the doping effects by shifting the Fermi level, which is reasonable, when the doping
level is low\cite{tt9,tt10,tt11}.
The semi-classic transport coefficients, such as  Seebeck coefficient S,  electrical conductivity with respect to scattering time  $\mathrm{\sigma/\tau}$ and  power factor with respect to scattering time $\mathrm{S^2\sigma/\tau}$,  as  a function of doping level  at room temperature using mBJ and mBJ+SOC are plotted in \autoref{t1}. Due to  electrical thermal conductivity $\kappa_e$=$L\sigma T$ (Lorenz number $L$=$\pi^2k_B^2/3e^2$, where $K_B$ is the Boltzmann constant, e is the charge of an electron.), the  electrical thermal conductivity has similar outlines with electrical conductivity.
The Fermi level moves into conduction bands, which means  n-type doping (negative doping levels) with the negative Seebeck coefficient.  The p-type doping (positive doping levels) with the positive Seebeck coefficient can be achieved by shifting Fermi level into the valence bands.
Although  the Seebeck
coefficient  is very large, when the Fermi level is in the middle of  band gap, the low electrical conductivity leads to very small power factor. Shifting the Fermi level into conduction bands or valence bands, the Seebeck coefficient (absolute value)
decreases, while  the electrical conductivity increases, which leads to  a  maximum  of power factor at certain doping level.

\begin{figure}
  \includegraphics[width=8cm]{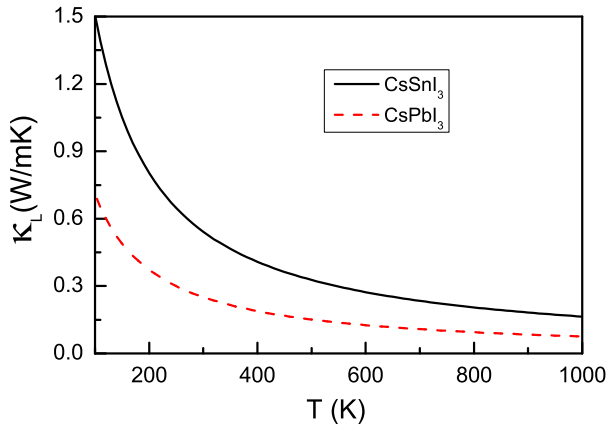}
  \caption{(Color online) The lattice  thermal conductivities $\kappa_L$ of $\mathrm{CsMI_3}$ (M=Sn and Pb)  as a function of temperature using GGA.}\label{t3}
\end{figure}

It has been proved that SOC has very important effects on power factor in many thermoelectric materials containing heavy elements\cite{so1,so2,so3,so4,so5,gsd3,gsd4,so6}. Calculated results show that SOC has obvious reduced influences on S and $\mathrm{\sigma/\tau}$ in n-type doping for both $\mathrm{CsSnI_3}$  and $\mathrm{CsPbI_3}$, but weak effects for p-type.
The large slope of density of states (DOS) near the energy band gap can induce a large Seebeck coefficient in narrow-gap semiconductors. This can be understood by the following formula: $S=\frac{\pi^2}{3}(\frac{k_B^2T}{e})[\frac{1}{n}\frac{dn(E)}{dE}+\frac{1}{\mu}\frac{d\mu(E)}{dE}]_{E=E_f}$\cite{z1},
where $n(E)$ and $\mu(E)$ are energy dependent carrier density and mobility, respectively.
It is found that the slope of DOS   using mBJ near the energy band gap is larger than that using mBJ+SOC  for conduction bands,
while they are nearly the same for valence bands. Band degeneracy, namely  band convergence, can induce lager slope of DOS. The SOC can reduce the slope of DOS   by removing band degeneracy. The SOC effects on energy bands can explain SOC influences on S.
The SOC-induced  reduced S and $\mathrm{\sigma/\tau}$ for n-type  lead to a very remarkable detrimental influence on power factor, especially for  $\mathrm{CsPbI_3}$.  However, the SOC has weak effects on p-type power factor.
For $\mathrm{CsPbI_3}$,  at the absence of SOC,  the best  n-type power factor  is larger than that in p-type doping. However, including SOC, it is opposite in considered doping range. Similar SOC-induced switch of best power factor between n-type and p-type can be achieved in  $\mathrm{Mg_2Sn}$ \cite{gsd3}.
 Therefore, including SOC is very important  in the theoretical prediction  of thermoelectric properties of $\mathrm{CsMI_3}$ (M=Sn and Pb).

\begin{figure*}
  \includegraphics[width=14cm]{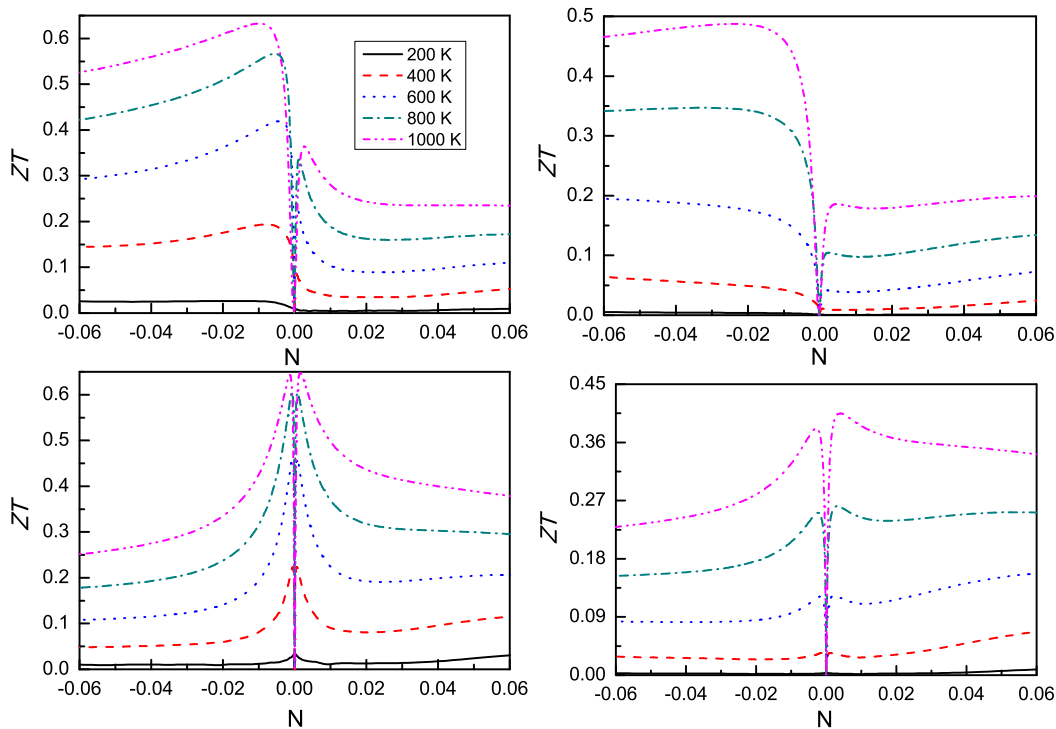}
  \caption{(Color online) The $ZT$ of $\mathrm{CsSnI_3}$ (Top panel) and $\mathrm{CsPbI_3}$ (Bottom panel) as a function of doping level with temperature  being 200, 400, 600 , 800 and 1000 (unit: K), and the scattering time $\mathrm{\tau}$  is 1 $\times$ $10^{-14}$ s (Left) and 1 $\times$ $10^{-15}$ s (Right). }\label{t4}
\end{figure*}

The power factor and  electronic thermal conductivity   with respect to scattering time  ($\mathrm{S^2\sigma/\tau}$ and $\mathrm{\kappa_e/\tau}$) of $\mathrm{CsSnI_3}$  and $\mathrm{CsPbI_3}$  as a function of doping level with temperature from 200 K to  1000 K using mBJ+SOC are shown in \autoref{t2}. For $\mathrm{CsSnI_3}$,  n-type doping has larger power factor than p-type doping, while p-type power factor is larger than n-type one for $\mathrm{CsPbI_3}$. if we assume scattering time  is constant, in considered doping and temperature range, the best power factor of $\mathrm{CsSnI_3}$ is nearly four times larger than one of  $\mathrm{CsPbI_3}$, and about two times larger for electronic thermal conductivity. The lattice thermal conductivity is important factor,  which significantly affects thermoelectric performance. The lattice  thermal conductivities $\kappa_L$ of $\mathrm{CsMI_3}$ (M=Sn and Pb) as a function of temperature   are shown in \autoref{t3}. Calculated results show ultralow lattice thermal conductivities in $\mathrm{CsMI_3}$ (M=Sn and Pb).  The lattice thermal conductivity can be assumed  to have weak dependence on doping level, and typically goes as 1/T.
The corresponding room-temperature lattice thermal conductivity is 0.54   $\mathrm{W m^{-1} K^{-1}}$ and 0.25 $\mathrm{W m^{-1} K^{-1}}$ for $\mathrm{CsSnI_3}$  and $\mathrm{CsPbI_3}$.  Theoretically, ultralow lattice thermal conductivities in many compounds have been predicted, such as $\mathrm{PbRbI_3}$ (0.10  $\mathrm{W m^{-1} K^{-1}}$), PbIBr (0.13  $\mathrm{W m^{-1} K^{-1}}$) and $\mathrm{K_2CdPb}$ (0.45  $\mathrm{W m^{-1} K^{-1}}$)\cite{ltc1}.
It is found that lattice thermal conductivity of $\mathrm{CsSnI_3}$
is nearly two times larger than one of  $\mathrm{CsPbI_3}$.

Due to the complexity of various carrier scattering mechanisms,  it is difficult to  calculate scattering time  $\tau$ from the first principles. To estimate thermoelectric conversion efficiency, the
thermoelectric figure of merit $ZT$ is calculated with hypothetical $\tau$=$10^{-14}$ and  $\tau$=$10^{-15}$ s, and are plotted
in \autoref{t4}.  The peak $ZT$ and  corresponding doping concentrations for both n- and p-type at 1000 K are summarized in \autoref{tab}. For $\mathrm{CsSnI_3}$, the n-type doping has larger $ZT$ than p-type doping, which is mainly due to the larger n-type Seebeck coefficient.  However, the $ZT$ between n- and p-type are nearly the same for  $\mathrm{CsPbI_3}$ due to almost the same  Seebeck coefficient. According to \autoref{t2} and \autoref{t3}, the total thermal conductivity $\mathrm{\kappa}$ is dominated by the lattice thermal conductivity $\mathrm{\kappa_L}$ in the very low doping level, but the electronic thermal conductivity $\mathrm{\kappa_e}$ becomes very larger than lattice thermal conductivity $\mathrm{\kappa_L}$  in slightly high doping region. These leads to very low doping concentration for peak $ZT$. Therefore, electronic thermal conductivity  of $\mathrm{CsMI_3}$ (M=Sn and Pb) is  a fatal disadvantage to gain more higher $ZT$ value.

\section{Discussions and Conclusion}
The CBM of  $\mathrm{CsMI_3}$ (M=Sn and Pb) is dominated by a giant spin-orbit coupling (SOC), especially for  $\mathrm{CsPbI_3}$. The SOC   removes the band degeneracy  of CBM by  spin-orbit splitting, which leads to  obvious effects on n-type Seebeck coefficient, and  further affects the power factor.  The  larger  spin-orbit splitting $\Delta$ leads to the more obvious detrimental influence on n-type power factor, which can be observed from \autoref{t2}. The similar SOC-induced detrimental influence on power factor has been observed  in $\mathrm{Mg_2Sn}$ and half-Heusler $\mathrm{ANiB}$ (A=Ti, Hf, Sc, Y; B=Sn, Sb, Bi)\cite{so2,gsd3}. Therefore, it is very important  for electronic part of  thermoelectric properties of $\mathrm{CsMI_3}$ (M=Sn and Pb)
to include SOC.

 $\mathrm{CsMI_3}$ (M=Sn and Pb) have been predicted to be three-dimensional topological insulators under reasonable hydrostatic pressure  using  a tight-binding analysis and first-principles calculations\cite{s15,s16}, which means that the electronic structures of $\mathrm{CsMI_3}$ (M=Sn and Pb) are easily tuned by pressure. Pressure-induced enhanced power factor has been predicted in $\mathrm{Mg_2Sn}$\cite{gsd3} and BiTeI\cite{gsd7} by the first-principles calculations.  Experimentally, it is possible to tune the thermoelectric properties of $\mathrm{CsMI_3}$ (M=Sn and Pb) by pressure.

In summary, mBJ and mBJ+SOC are  chosen to investigate  electronic structures and electronic part of thermoelectric properties  of halide perovskites $\mathrm{CsMI_3}$ (M=Sn and Pb).  The strength of  SOC influences on  CBM  is very large, especially for $\mathrm{CsPbI_3}$, which gives rise to obvious detrimental influence on n-type power factor.
The lattice thermal conductivities of $\mathrm{CsMI_3}$ (M=Sn and Pb) are performed with GGA, and ultralow lattice thermal conductivities are predicted, which is very key for providing high thermoelectric performance. At 1000 K, in low doping level,  the figure of merit $ZT$ is up to about 0.6 with $\tau$=$10^{-14}$, and about 0.4 with $\tau$=$10^{-15}$.
The present work provides a platform to search potential thermoelectric materials from perovskite compounds.

\begin{acknowledgments}
This work is supported by the National Natural Science Foundation of China (Grant No. 11404391). We are grateful to the Advanced Analysis and Computation Center of CUMT for the award of CPU hours to accomplish this work.
\end{acknowledgments}

\end{document}